\documentclass[12pt]{article}
\usepackage{amsfonts}
\usepackage{graphicx}
\usepackage{amssymb} 
\usepackage{amscd}
\usepackage{amsmath}
\begin{document}
\renewcommand{\thefootnote}{\fnsymbol{footnote}}
\begin{titlepage}

\vspace{10mm}
\begin{center}
{\Large\bf A possible method for non-Hermitian and non-$PT$-symmetric Hamiltonian systems}

\vspace{16mm}

{{\large Jun-Qing Li${}^{1}$, Yan-Gang Miao${}^{1,2,3,}$\footnote{{Corresponding author.
E-mail address: miaoyg@nankai.edu.cn}}, and Zhao Xue${}^{1}$}\\

\vspace{6mm}

${}^{1}${\normalsize \em School of Physics, Nankai University, Tianjin 300071, China}

\vspace{3mm}
${}^{2}${\normalsize \em Kavli Institute for Theoretical Physics China, CAS, Beijing 100190, China}

\vspace{3mm}
${}^{3}${\normalsize \em Bethe Center for Theoretical Physics and Institute of Physics, University of Bonn, \\
Nussallee 12, D-53115 Bonn, Germany}}

\end{center}

\vspace{10mm}
\centerline{{\bf{Abstract}}}
\vspace{6mm}
\noindent
A possible method to investigate non-Hermitian Hamiltonians is suggested through finding a Hermitian operator $\eta_+$ and defining the annihilation and creation operators to be $\eta_+$-pseudo-Hermitian
adjoint to each other. The operator $\eta_+$ represents the $\eta_+$-pseudo-Hermiticity of Hamiltonians. As an example, a non-Hermitian and non-$PT$-symmetric Hamiltonian with imaginary linear coordinate  and linear momentum terms is constructed and analyzed in detail. The operator $\eta_+$  is found, based on which, a real spectrum and a positive-definite inner product, together with the probability explanation of wave functions, the orthogonality of eigenstates, and the unitarity of time evolution, are obtained for the non-Hermitian and non-$PT$-symmetric Hamiltonian. Moreover, this Hamiltonian turns out to be coupled when it is extended to the canonical noncommutative
space with noncommutative spatial coordinate operators and noncommutative momentum operators
as well. Our method is applicable to the coupled Hamiltonian. Then the first and second order noncommutative corrections of energy levels are calculated, and in particular the reality of energy spectra, the positive-definiteness of inner products, and the related properties (the probability explanation of wave functions, the orthogonality of eigenstates, and the unitarity of time evolution) are found not to be altered by the noncommutativity.


\vspace{3mm}

\end{titlepage}

\newpage
\renewcommand{\thefootnote}{\arabic{footnote}}
\setcounter{footnote}{0}
\setcounter{page}{2}

\section{Introduction}
A non-Hermitian Hamiltonian with a complex potential usually has complex
eigenvalues  and such a system
does not maintain the conservation of probability. However, the non-Hermitian Hamiltonian with a class of quasi-Hermiticity was proposed~\cite{SGH}
in which the real eigenvalues and the conservation of probability are possible.
Recently  the real eigenvalues and corresponding eigenstates of a non-Hermitian Hamiltonian associated with some symmetry
have been paid more attention to, such as the $\eta$-pseudo-Hermitian Hamiltonian~\cite{PAULI,Ali1,Ali2,Ali3} and the $PT$-symmetric Hamiltonian~\cite{BBR}. The former satisfies 
\begin{equation}
H=  \eta^{-1}H^\dagger \eta,\label{pseudo}
\end{equation}
where the invertible operator $\eta$ is linear Hermitian,
and  the latter satisfies the $PT$ symmetry, $H=  \left(PT\right)^{-1}H \left(PT\right)$, where $P$ and $T$ stand for the parity and  time reversal transformations, respectively. In addition,
some experiments~\cite{HTRD,YLA} on $PT$-symmetric ($P$-pseudo-Hermitian) Hamiltonians have been carried out in the region of optics.

Roughly speaking, there exist two methods that are used to study an ordinary  (Hermitian) Hamiltonian system.
The usual one focuses on solving the Schr\"odinger equation under certain boundary conditions in order to calculate eigenvalues and eigenstates.
The other method, which
is useful for dealing with the systems such as the harmonic oscillator,
is associated closely with annihilation and creation operators and their commutation relations.
Nevertheless,
in quantum theories on non-Hermitian Hamiltonians, the former method is commonly adopted in literature,
see, for instance, the review article~\cite{CBM}, while the latter cannot be utilized directly because imaginary terms appear in a non-Hermitian Hamiltonian.
In this paper we give a possible method through redefining annihilation and creation operators, and as an  application beyond the non-Hermitian $PT$-symmetric quantum theory~\cite{BBR}, we at first
construct a non-Hermitian and non-$PT$-symmetric Hamiltonian that is decoupled, 
analyze its spectrum and inner product,
and then apply our method to a coupled Hamiltonian that is given by extending  the decoupled one to the canonical noncommutative space.
Our method is therefore complementary to the non-Hermitian $PT$-symmetric method.

This paper is organized as follows. In the next section,
we elaborate our method for an $\eta_+$-pseudo-Hermitian Hamiltonian\footnote{In general,
$\eta_+$ does not coincide with $\eta$.
The subscript ``+'' means that $\eta_+$ is associated with a positive-definite inner product.}
by redefining annihilation and creation operators
that are $\eta_+$-pseudo-Hermitian (no longer Hermitian) adjoint to each other. The key point of this method
is to find out an $\eta_+$ operator that represents an inherent symmetry of the non-Hermitian Hamiltonian, {\em i.e.}, the $\eta_+$-pseudo-Hermiticity, also called the $\eta_+$-pseudo-Hermitian self-adjoint condition. With such an $\eta_+$, one can deduce that the non-Hermitian Hamiltonian
has a real spectrum with lower boundedness and a positive-definite inner product.\footnote{Here we have to mention an earlier work~\cite{GR} which also dealt with a non-Hermitian
Hamiltonian system.
Although the real spectrum was given there, the non-Hermiticity was not properly treated and more severely
the positive-definiteness of inner products was completely ignored. In fact, the annihilation and creation operators defined in ref.~\cite{GR} are no longer Hermitian adjoint to each other, which gives rise to the incorrect treatment to that non-Hermitian Hamiltonian.}
In section 3, as an application beyond the non-Hermitian $PT$-symmetric quantum theory~\cite{BBR}, we construct
a non-Hermitian and non-$PT$-symmetric Hamiltonian that is decoupled, and find out $\eta_+=PV$ through introducing a special  operator $V$.
Then we redefine the
annihilation and creation operators in such a way that they are $PV$-pseudo-Hermitian adjoint to each other, and as expected, obtain a real energy spectrum and a positive-definite inner product. 
In section 4, the decoupled non-Hermitian and non-$PT$-symmetric Hamiltonian is extended to the canonical
noncommutative phase space with noncommutative coordinate operators and noncommutative momentum operators as well, and
it turns out to be coupled. Our method is applicable to the coupled Hamiltonian.
We then calculate the noncommutative corrections of energy levels
up to the first and second orders in noncommutative parameters, respectively, and in particular we give an interesting result that the reality of energy spectra
and the positive-definiteness of inner products are not altered by the noncommutativity of phase space.
Finally, we make a conclusion in section 5.

\section{The method for $\eta_+$-pseudo-Hermitian systems}

At first, we review the modified inner product. The condition that an
$\eta_+$-pseudo-Hermitian observable ${\cal A}$ should obey takes the following form~\cite{PAULI,Ali3,ALICPT}, {\em i.e.},
the $\eta_+$-pseudo-Hermitian self-adjoint condition,
\begin{eqnarray}
{\cal A}^{\ddagger}\equiv \eta^{-1}_+ {\cal A}^{\dagger} \eta_+ ={\cal A},\label{observable}
\end{eqnarray}
where the superscript ``$\ddagger$" stands for the action of an $\eta_+$-pseudo-Hermitian adjoint to an operator. Note that the $\eta_+$-pseudo-Hermitian adjoint becomes the ordinary Hermitian adjoint when $\eta_+$ takes the identity operator, in which case the $\eta_+$-pseudo-Hermitian Hamiltonian turns back to the Hermitian Hamiltonian. 
Related to eq.~(\ref{observable}), the definition of the modified bra vector states
is as follows:
\begin{eqnarray}
{}^\ddagger\langle\varphi(x)|\equiv \langle\varphi(x)|\eta_+ ,\label{stateadjoint}
\end{eqnarray}
where $\langle\varphi(x)|$ denotes a usual bra vector state that is
Hermitian adjoint to a (usual) ket vector state $|\varphi(x)\rangle$, {\em i.e.},
$\langle\varphi(x)|=(|\varphi(x)\rangle)^{\dagger}$.
It is convenient to use the notation of the left hand side of eq.~(\ref{stateadjoint}), {\em i.e.}, 
the notation with hidden $\eta_+$, 
which will be seen evidently in the following context.
The modified bra vector state may be called the $\eta_+$-pseudo-Hermitian adjoint to
the ket vector state and it becomes the normal one in the Hermitian quantum mechanics where $\eta_+$ is just the identity operator.
Therefore, the modified inner product in the Hilbert space for an $\eta_+$-pseudo-Hermitian Hamiltonian system naturally has the form,
\begin{eqnarray}
{}^\ddagger\langle\varphi(x)|\psi(x)\rangle=\langle\varphi(x)|\eta_+|\psi(x)\rangle,\label{innerpro}
\end{eqnarray}
which can be understood as a generalized inner product. Since $\eta$ (see eq.~(\ref{pseudo})) was called  by Pauli~\cite{PAULI} the indefinite metric
in the Hilbert space, $\eta_+$ is then called the positive-definite metric because it gives rise to~\cite{SGH,ML} a real and
positive-definite norm or probability,
$\langle\psi(x)|\eta_+|\psi(x)\rangle \geq 0$. Note that the operator
$\eta_+$ is in general required~\cite{PAULI,Ali1,Ali2,Ali3}
to be linear Hermitian\footnote{In our recent work~\cite{ML} we discuss the anti-linear anti-Hermitian case and obtain some interesting results.}
and invertible,
which ensures not only the reality of the average of physical observables but also
the reality and positivity of the probability.
In addition, we mention that 
the self-adjoint condition (eq.~(\ref{observable})) is consistent with the
modified inner product (see eq.~(\ref{innerpro})),
{\em i.e.}, ${}^\ddagger\langle {\cal A} \varphi(x)|\psi(x)\rangle \equiv \langle\varphi(x)| {\cal A}^{\dagger} \eta_+|\psi(x)\rangle= \langle\varphi(x)| \eta_+\left(\eta_+^{-1}{\cal A}^{\dagger} \eta_+\right)|\psi(x)\rangle= \langle\varphi(x)| \eta_+{\cal A}^{\ddagger}|\psi(x)\rangle= \langle\varphi(x)| \eta_+{\cal A}|\psi(x)\rangle \equiv {}^\ddagger\langle \varphi(x)|{\cal A} \psi(x)\rangle$. 
We point out that it is the
requirement of positive norms that makes it a quite nontrivial task to find out the metric  $\eta_+$ even for a simple
non-Hermitian Hamiltonian,
which can be seen clearly from our non-Hermitian and non-$PT$-symmetric models in the two sections below.

Next, we redefine the
creation operator as the $\eta_+$-pseudo-Hermitian adjoint to a particularly chosen annihilation
operator (for concrete procedures see the following two sections) as follows:
\begin{equation}
a^\ddagger\equiv\eta^{-1}_+ a^\dagger \eta_+,\label{ceation}
\end{equation}
which is quite different from the definition in the Hermitian quantum mechanics.
Note that the redefined creation and annihilation operators are $\eta_+$-pseudo-Hermitian adjoint to each other, that is, we
have\footnote{It is easy to verify this equality, that is,
$(a^\ddagger)^\ddagger = \eta_+^{-1} \left(\eta^{-1}_+ a^\dagger \eta_+\right)^\dagger \eta_+ =\eta^{-1}_+ \eta_+ a \eta^{-1}_+ \eta_+ =a$, where
the Hermiticity of $\eta_+$ has been utilized.}
$a=(a^\ddagger)^\ddagger$.
In addition, we can prove that $a$ and $a^\ddagger$ are $\eta_+$-pseudo-Hermitian adjoint to each other
with respect to the generalized inner product
eq.~(\ref{innerpro}), that is, considering eqs.~(\ref{innerpro}) and (\ref{ceation}) we get\footnote{We can verify  ${}^\ddagger\langle a \varphi(x)|\psi(x)\rangle= \langle\varphi(x)|a^\dagger\eta_+|\psi(x)\rangle =
\langle\varphi(x)|\eta_+\left(\eta_+^{-1}a^\dagger\eta_+\right)|\psi(x)\rangle = \langle\varphi(x)|\eta_+a^\ddagger|\psi(x)\rangle= {}^\ddagger\langle \varphi(x)|a^\ddagger \psi(x)\rangle$.}
\begin{eqnarray}
{}^\ddagger\langle a \varphi(x)|\psi(x)\rangle= {}^\ddagger\langle \varphi(x)|a^\ddagger \psi(x)\rangle. \label{apseudohermitian}
\end{eqnarray}
This shows that the redefinition of the annihilation and creation operators is consistent with the definition of
the modified bra vector states eq.~(\ref{stateadjoint}).
The formula $a=(a^\ddagger)^\ddagger$ becomes the one we are quite familiar with, {\em i.e.},  $a=(a^\dagger)^\dagger$,
when $\eta_+$ takes the identity operator, {\em i.e.}, when
an $\eta_+$-pseudo-Hermitian system becomes a Hermitian one.
Considering the well-known commutation relations satisfied by the usual annihilation and creation operators in the conventional quantum mechanics,
we require that the redefined annihilation and creation operators in the $\eta_+$-pseudo-Hermitian quantum mechanics comply with
\begin{eqnarray}
[a,a^{\ddag}]=1, \qquad
[a,a]=0=[a^{\ddag},a^{\ddag}],\label{acalgrbra}
\end{eqnarray}
which turns consistently back to the usual commutation relations when $\eta_+$ becomes the identity operator, {\em i.e.}, that $\eta_+$ becomes the identity operator is equivalent to that the $\eta_+$-pseudo-Hermitian self-adjoint becomes the Hermitian self-adjoint, ${\ddag} \rightarrow \dag$.

At last, we define the corresponding number operator in the pseudo-Hermitian quantum mechanics as follows:
\begin{equation}
N\equiv a^\ddagger a,\label{number}
\end{equation}
which, as a physical observable, is of course $\eta_+$-pseudo-Hermitian self-adjoint,\footnote{Considering eqs.~(\ref{ceation}) and (\ref{number}),
we have
$N^\ddagger=\left(\eta^{-1}_+ a^\dagger \eta_+ a\right)^{\ddagger}=\eta^{-1}_+ \left(\eta^{-1}_+ a^\dagger \eta_+ a\right)^{\dagger} \eta_+
=\eta^{-1}_+ a^\dagger \eta_+ a \eta^{-1}_+ \eta_+ =a^\ddagger a =N$, where the Hermiticity of $\eta_+$ has been used.} 
{\em i.e.}, $N^\ddagger=N$. More precisely, the number operator $N$ is self-adjoint with respect to the
generalized inner product,\footnote{Using eq.~(\ref{innerpro}), we can verify ${}^\ddagger\langle N \varphi(x)|\psi(x)\rangle= \langle\varphi(x)|N^\dagger\eta_+|\psi(x)\rangle =
\langle\varphi(x)|\eta_+\left(\eta_+^{-1}N^\dagger\eta_+\right)|\psi(x)\rangle = \langle\varphi(x)|\eta_+N^\ddagger|\psi(x)\rangle= {}^\ddagger\langle \varphi(x)|N \psi(x)\rangle$.}
\begin{eqnarray}
& &{}^\ddagger\langle N \varphi(x)|\psi(x)\rangle= {}^\ddagger\langle \varphi(x)|N \psi(x)\rangle,\label{Npseudohermitian}
\end{eqnarray}
which shows that the definition of modified bra and ket vector states is consistent with the self-adjoint requirement of physical observables.
Using eqs.~(\ref{acalgrbra}) and (\ref{number}), we can verify the following commutation relations in the pseudo-Hermitian quantum mechanics,
\begin{eqnarray}
[N,a^{\ddagger}]=a^{\ddagger}, \qquad [N,a]=-a.\label{Nacalgrbra}
\end{eqnarray}
Consequenly, we provide a possible method for an $\eta_+$-pseudo-Hermitian system. The remaining task is just to deduce some useful
formulae, such as the $n$-particle state and the ladder property of redefined creation and annihilation operators, which is fulfilled in Appendix A for the completeness of our method.

In addition, we note that the unitarity of time evolution is guaranteed with respect to the modified inner product (eq.~(\ref{innerpro}))
in the $\eta_+$-pseudo-Hermitian quantum mechanics.
Considering the $\eta_+$-pseudo-Hermitian self-adjoint of the Hamiltonian, {\em i.e.}, $H=  \eta_+^{-1}H^\dagger \eta_+$, and the time evolution of an
initial state $|\psi(0)\rangle$, $|\psi(t)\rangle=e^{-iHt}|\psi(0)\rangle$, we have\footnote{It is obvious to prove the relation: ${}^\ddagger \langle \psi(t)|\psi(t)\rangle  \equiv  \langle \psi(t)|\eta_+ |\psi(t)\rangle
= \langle \psi(0)|e^{+iH^{\dagger}t}\eta_+ e^{-iHt}|\psi(0)\rangle 
= \langle \psi(0)|\eta_+ (\eta_+^{-1}e^{+iH^{\dagger}t}\eta_+) e^{-iHt}|\psi(0)\rangle
= \langle \psi(0)|\eta_+ (e^{+iHt}) e^{-iHt}|\psi(0)\rangle
= \langle \psi(0)|\eta_+ |\psi(0)\rangle
 \equiv  {}^\ddagger \langle \psi(0)|\psi(0)\rangle$.}
\begin{eqnarray}
{}^\ddagger \langle \psi(t)|\psi(t)\rangle 
=  {}^\ddagger \langle \psi(0)|\psi(0)\rangle, \label{unitarity}
\end{eqnarray}
which gives the unitary time evolution.

As a summary, we point out that the characteristic of our method is to adopt the orthonormal basis of Hamiltonians.
Although the biorthonormal basis~\cite{Wong} has been applied~\cite{Ali1,Ali2,Ali3} to pseudo-Hermitian Hamiltonian systems,
it is interesting to investigate whether the usual treatment (to consider just the orthonormal basis of Hamiltonians but not that of
the Hermitian conjugate of Hamiltonians) is still available to such non-Hermitian systems.
Our way to realize this goal is to find out the specific operator $\eta_+$ and then to
make the redefinition of annihilation and creation operators (eq.~(\ref{ceation}))
that are adjoint to each other with respect to the generalized inner product (eq.~(\ref{innerpro})). For the concrete
procedure, see the next two sections. We emphasize that the operator $\eta_+$ is in general nontrivial, that is, it is impossible to
reduce $\eta_+$ to be the identity through a basis transformation. The reason is that the Hamiltonian of an $\eta_+$-pseudo-Hermitian system
is no longer Hermitian self-adjoint with respect to the usual inner product:
$\langle H \varphi(x)|\psi(x)\rangle \neq \langle \varphi(x)|H \psi(x)\rangle$,
but $\eta_+$-pseudo-Hermitian self-adjoint with respect to the generalized inner product
(see eqs.~(\ref{stateadjoint}) and (\ref{innerpro})):
${}^\ddagger\langle H \varphi(x)|\psi(x)\rangle= {}^\ddagger\langle \varphi(x)|H \psi(x)\rangle$. 
The operator $\eta_+$, as an inherent symmetry of $H$,  
can never be eliminated through a basis transformations. Finally, we make a comment that our method 
may be understood as a generalized Fock space representation for non-Hermitian and non-$PT$-symmetric quantum systems due to the existence of a nontrivial $\eta_+$, and that it  can be applied to deal with such 
systems mainly by the
redefinitions of annihilation, creation, and number operators and by the reconstruction of their commutation relations.  

\section{The non-Hermitian and non-$PT$-symmetric system}
In this section we investigate a concrete non-Hermitian Hamiltonian by means of the 
method provided in the above section. 
In order to show that our method is complementary to the non-Hermitian $PT$-symmetric method~\cite{BBR}, we construct a non-Hermitian and non-$PT$-symmetric Hamiltonian. We add two imaginary terms which are proportional to $i({x_1}+{x_2})$ and $i({p_1}+{p_2})$, respectively, 
to the Hamiltonian of an isotropic planar
oscillator, and then give a new Hamiltonian:
\begin{equation}
H=\frac1 2\left(p_1^2+x_1^2\right)+\frac1 2\left(p_2^2+x_2^2\right)+i\left\{A\left(x_1+x_2\right)+B\left({p_1}+{p_2}\right)\right\},\label{H1}
\end{equation}
where $A$ and $B$ are real parameters; $x_j$ and $p_j$ ($j=1,2$) are two pairs of canonical coordinate operators and their
conjugate momentum operators, they are Hermitian and satisfy the standard Heisenberg commutation relations, 
where $\hbar$ is set be unity through out this paper. 
This Hamiltonian is decoupled,
and normally it is enough for us to analyze
its one-dimensional part in this section.
However, we shall see that this is a good enough model for us to illustrate our method clearly, and that 
it is quite nontrivial to find out operator $\eta_+$ for such a simple model. In particular,
the decoupled Hamiltonian will turn out to be coupled 
when it is extended to the noncommutative space in the next section, and our method is still applicable to the coupled Hamiltonian. That is the reason why we 
choose such a decoupled two-dimensional Hamiltonian in this section.

The Hamiltonian eq.~(\ref{H1}) is obviously non-Hermitian and non-$PT$-symmetric due to the different properties of $ix$ and $ip$ under the $PT$ transformation~\cite{BBR}. For a non-Hermitian {\em but} $PT$-symmetric Hamiltonian, there exists a well-established theory called $PT$-symmetric quantum mechanics~\cite{CBM}. In order to extend our discussion beyond the $PT$-symmetric theory, we particularly construct the non-Hermitian and non-$PT$-symmetric Hamiltonian depicted by  eq.~(\ref{H1}),  convert it into an $\eta_+$-pseudo-Hermitian Hamiltonian, and then deal with it by using the method demonstrated in the above section.
 
We notice that the decoupled Hamiltonian can easily be diagonalized and rewritten as
\begin{equation}
H=H_1+H_2+(A^2+B^2),\label{H2}
\end{equation}
where the new variables are defined by
\begin{eqnarray}
H_1\equiv \frac12(P^2_1+X^2_1), && H_2\equiv \frac12(P^2_2+X^2_2),\nonumber\\
P_1\equiv p_1+iB, && X_1\equiv x_1+iA,\nonumber\\
P_2\equiv p_2+iB, && X_2\equiv x_2+iA .\label{NewV}
\end{eqnarray}
Eq.~(\ref{H2}), together with eq.~(\ref{NewV}), looks like the usual Hamiltonian of a decoupled two-dimensional  harmonic
oscillator, but in fact, it is not the case. Now $X_j$ and $P_j$,
though satisfying the standard Heisenberg  commutation relations, 
\begin{equation}
[X_j,P_k]=i{\delta}_{jk}, \qquad [X_j,X_k]=0=[P_j,P_k], \qquad
j,k=1,2,\label{cr2}
\end{equation}
are not Hermitian self-adjoint, but as expected, they are $\eta_+$-pseudo-Hermitian self-adjoint and have real average values with respect to the generalized inner product (cf. eq.~(\ref{innerpro})), and so are  $H_1$ and $H_2$.

Now we begin the investigation of the system governed by the Hamiltonian eq.~(\ref{H1}) or eq.~(\ref{H2}).
It is subtle to find $\eta_+$ for the system.

We give a two-step way for the construction of the desired operator $\eta_+$.
In the first step, inspired by Lee and Wick~\cite{LW}, we define
operator $V$ as follows,
\begin{equation}
V\equiv V_1V_2,\qquad V_1\equiv (-1)^{H_1-\frac{1}{2}}, \qquad V_2\equiv (-1)^{H_2-\frac{1}{2}}.\label{V}
\end{equation}
$V$, $V_1$, and $V_2$ are invertible.
Note that the exponential factor $-\frac{1}{2}$ in $V_1$ and $V_2$ is introduced in order to eliminate the zero-point energy in $H_1$ and $H_2$ and then to express $V_1$ and $V_2$ in terms of number operators (cf. eq.~(\ref{HN})).  
Here $V$ is $P$-pseudo-Hermitian self-adjoint (see Appendix B for the proof),
\begin{equation}
V=P^{-1}V^{\dag}P,\label{VPseudo}
\end{equation}
which is different from the case in ref.~\cite{LW} where a Hermitian
operator was introduced.
Moreover, we point out that $V$ is defined in terms of the Hamiltonian, which is more intuitive than the definition of the operator $C$~\cite{BBR}  for positive-definite inner products in the
$PT$-symmetric quantum mechanics, where $C$ is defined by unknown eigenstates of a non-Hermitian {\em but} $PT$-symmetric Hamiltonian.
Then, in the second step we set $\eta_+$ be the product of $P$ and $V$,
\begin{equation}
\eta_+=PV,\label{eta}
\end{equation}
which is linear Hermitian and invertible. Note that $V$ is linear non-Hermitian because $H_1$ and $H_2$ are non-Hermitian.
It is easy to prove the Hermiticity of $\eta_+$ by considering
the $P$-pseudo-Hermitian self-adjoint of $V$ (eq.~(\ref{VPseudo})), that is,
$\eta_+^\dagger=V^\dagger P^\dagger=P (P^{-1}V^\dagger P)=P V=\eta_+$. 
In addition, because $\eta_+$ is related to the Hamiltonian through $V$ the generalized  inner product with respect to this $\eta_+$
(see eq.~(\ref{innerpro}))
can be called a dynamical inner product as the $CPT$ inner product~\cite{BBR} was called.

Using eqs.~(\ref{NewV})-(\ref{V}), we deduce the $PV$-pseudo-Hermiticity of $X_j$ and $P_j$,
\begin{eqnarray}
X_j^{\ddagger} &\equiv & \left(PV\right)^{-1}X_j^{\dag}\left(PV\right)= X_j, \nonumber \\
P_j^{\ddagger} &\equiv & \left(PV\right)^{-1}P_j^{\dag}\left(PV\right)= P_j,\label{XPetaP}
\end{eqnarray}
which is proved in Appendix C. 
If $a_j$ is specifically chosen as $a_j=\frac{1}{\sqrt{2}}(X_j+iP_j)$, $j=1,2$, 
we have from eq.~(\ref{ceation}) the operator $a_j^{\ddagger}$ as the $PV$-pseudo-Hermitian adjoint to the operator $a_j$, $a_j^{\ddagger}\equiv(PV)^{-1}a^{\dagger}_j(PV)=\frac{1}{\sqrt{2}}(X_j-iP_j)$, $j=1, 2$.
This shows the subtleness of the construction of  $\eta_+=PV$ because we can then 
rewrite the Hamiltonian eq.~(\ref{H1}) in terms of  $a_j^{\ddagger}$ and $a_j$ that can be verified to satisfy the basic requirement in terms of eq.~(\ref{cr2}), $[a_j,a^{\ddag}_k]=\delta_{jk}$, 
$[a_j,a_k]=0=[a^{\ddag}_j,a^{\ddag}_k]$, $j, k=1,2$. In accordance with the formulae given in Appendix A, we can now write the number operator which is $PV$-pseudo-Hermitian self-adjoint,\footnote{Repeated subscripts do not sum except for extra indications.} $N_j=a_j^{\ddagger}a_j$, $j=1,2$, 
and get the expected commutation relations by using the algebraic relations of $a_j$ and $a_j^{\ddagger}$ and the expression of the number operator,
$[N_j,a^{\ddagger}_k]=a^{\ddagger}_j\delta_{jk}$, $[N_j,a_k]=-a_j\delta_{jk}$, $j,k=1,2$.
Furthermore, given $|n_j\rangle$ a set of eigenstates of the number operator $N_j$, {\em i.e.},
$N_j|n_j\rangle=n_j|n_j\rangle$, $j=1,2$,
if its inner product defined by eq.~(\ref{innerpro})
is positive definite, $a^{\ddagger}_j$ and $a_j$ can finally be convinced to be
the creation and annihilation operators that satisfy the property of ladder operators,
$a^{\ddagger}_j|n_j\rangle = \sqrt{n_j+1}\,|n_j+1\rangle$, $a_j|n_j\rangle =\sqrt{n_j}\,|n_j-1\rangle$, $j=1,2$,
and $H_1$ and $H_2$ can be expressed in terms of the number operators as follows:
\begin{eqnarray}
H_1=N_1+\frac{1}{2},\qquad
H_2=N_2+\frac{1}{2}.\label{HN}
\end{eqnarray}

At present we turn to the verification 
that the generalized inner product defined by eq.~(\ref{innerpro}) is positive definite
for our choice $\eta_+=PV$. Due to eq.~(\ref{Naverage}) (in Appendix A), we only need to prove the normalization of the ground state,
$\langle0|PV|0\rangle =1$. Utilizing eqs.~(\ref{V}) and (\ref{HN}) together with the property of the number operator,
we have $V|0\rangle =(-1)^{H_1+H_2-1} |0\rangle=(-1)^{N_1+N_2} |0\rangle=|0\rangle$, and thus obtain $\langle0|PV|0\rangle =\langle0|P|0\rangle$.
In addition, considering the wavefunction of the ground state,
$\varphi_0(X_j)= \frac{1}{\sqrt[4]{\pi}}\exp(-\frac{1}{2}X_j^2+BX_j)$,
where $j=1,2$, 
together with the definitions of $X_j$ and $P_j$ (see eq.~(\ref{NewV})), we can calculate the generalized inner product with respect to the ground state in terms of the Cauchy's residue theorem of
the complex function theory (see Figure 1 for the details),
\begin{eqnarray}
\langle0|PV|0\rangle &=&\langle0|P|0\rangle=\int^{+\infty+iA}_{-\infty+iA} \overline{\varphi}_0(X_j)\,P\, \varphi_0(X_j)\,dX_j \nonumber \\
&=&\frac{1}{\sqrt{\pi}}\int^{+\infty+iA}_{-\infty+iA}\exp\left[-\left(x_j-iA\right)^2\right]d\left(x_j+iA\right)\nonumber \\
&=&1,\qquad j=1,2,\label{PVinner}
\end{eqnarray}
where $\overline{\varphi}_0$ denotes the complex conjugate to ${\varphi}_0$.
Note that the symbols $X_j$ and $x_j$ in the above equation no longer stand for operators but coordinates.
Eq.~(\ref{PVinner}) definitely gives the normalization of the ground state, which, together with eq.~(\ref{Naverage}),
leads to $\langle n_j|PV|n_j \rangle =1$, where $j=1, 2$. That is, we at last
prove 
the positive definiteness of the generalized inner product (defined by eq.~(\ref{innerpro})) for the set of eigenstates of the number operator, $|n_j\rangle$.
\begin{figure}[!htbp]
\centering
\includegraphics[width=0.8\textwidth,height=6cm]{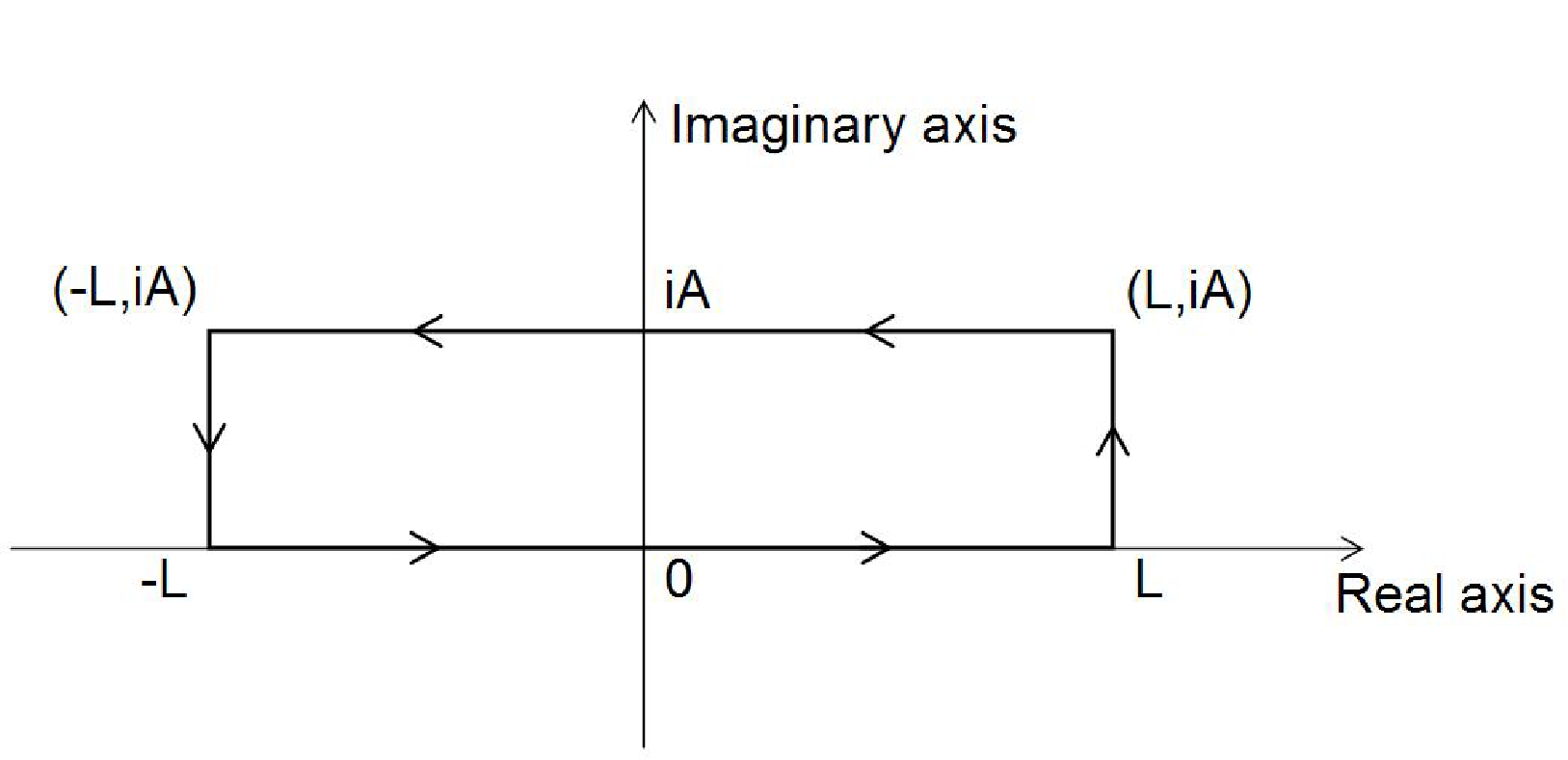}
\caption{\small We choose the rectangle with length $L$ and width $A$ in the complex plane as the contour. Inside the rectangle the integrand
$\overline{\varphi}_0(x+iy)\,P\, \varphi_0(x+iy)$ is analytic, {\em i.e.}, no singular points exist,
thus the contour integration is zero by means of the Cauchy's residue theorem in a simply connected domain. The integrations along the two perpendicular
(right and left) sides equal
$\exp(-L^2)\int_{0}^{A}\exp(y^2)\left[-\sin(2Ly)\pm i\cos(2Ly)\right]dy$,
they are also vanishing when the length $L$ tends to infinity. In consequence, the integrations on the top and bottom sides
of the rectangle equal if along the same direction and in the limit $L\rightarrow \infty$, as given explicitly by eq.~(\ref{PVinner}).}
\label{figure:dctdft}
\end{figure}

Alternatively, we can exactly solve the wavefunctions for any excited states related with the Hamiltonian
eq.~(\ref{H1}) or eq.~(\ref{H2}):
\begin{eqnarray}
\varphi_{n_1n_2}(X_1,X_2)=\varphi_{n_1}(X_1)\varphi_{n_2}(X_2),\label{ES}
\end{eqnarray}
where
\begin{eqnarray}
\varphi_{n_j}(X_j)=\frac{1}{\sqrt[4]{\pi}}(2^{n_j} {n_j}!)^{-\frac{1}{2}}\;e^{-\frac{1}{2}X_j^2+BX_j}\,H_{n_j}(X_j),  \qquad j=1,2,\label{ES1}
\end{eqnarray}
and $H_{n_j}(X_j)$ denotes the Hermite polynomial of the $n_j$-th degree with the argument $X_j$. 
Therefore, by considering $V\varphi_{n_j}(X_j)=(-1)^{N_1+N_2}\varphi_{n_j}(X_j)=(-1)^{n_j}\varphi_{n_j}(X_j)$ and using the same contour as in Figure 1
we can prove that the inner product for oscillator ${}^{\sharp} 1$ or oscillator ${}^{\sharp} 2$ is orthogonal and normalized, {\em i.e.},
\begin{eqnarray}
\langle n_j|PV|m_j\rangle =\int^{+\infty+iA}_{-\infty+iA} \overline{\varphi}_{n_j}(X_j)\,PV\, \varphi_{m_j}(X_j)\,dX_j=\delta_{n_j m_j}, \qquad j=1,2.
\label{PVinner1}
\end{eqnarray}
This shows from an alternative point of view that the positive definiteness of the inner product is guaranteed.

As a result, using eq.~(\ref{HN})
we can easily rewrite the Hamiltonian eq.~(\ref{H2}) in terms of the number operators as follows:
\begin{equation}
H=(N_1 +N_2 +1)+(A^2+B^2),\label{H3}
\end{equation}
and then give its real and positive spectrum,
\begin{eqnarray}
E_{n_1n_2}=(n_1+n_2 +1)+(A^2+ B^2), \label{EP}
\end{eqnarray}
where $n_1, n_2=0,1,2,\cdots$.
In an alternative way, we can get the same spectrum eq.~(\ref{EP}) if the Hamiltonian eq.~(\ref{H3}) acts
directly on the eigenfunction eq.~(\ref{ES}).

At the end of this section we point out that the non-Hermitian and non-$PT$-symmetric Hamiltonian (eq.~(\ref{H1})) has the inherent $PV$-pseudo-Hermiticity, 
\begin{equation}
H^{\ddagger}\equiv (PV)^{-1}H^{\dagger}(PV)=H,
\end{equation}
which was unknown initially {\em but} is exposed later by the finding of the operator $\eta_+=PV$. This property is quite obvious when we verify it by using eqs.~(\ref{H2})-(\ref{V}), that is,
$(PV)^{-1}H^{\dagger}(PV)=V^{-1}(P^{-1}H^{\dagger}P)V=V^{-1}HV=H$, where $[H,V]=0$ is used in the last equality.
In addition, we emphasize that $X_j$ and $P_j$ also have $PV$-pseudo-Hermiticity, see eq.~(\ref{XPetaP}), 
and thus they, rather than the Hermitian operators $x_j$ and $p_j$,  are physical observables with real average values under the generalized definition of inner products (cf. eq.~(\ref{innerpro})).
We conclude that this $PV$-pseudo-Hermitian Hamiltonian, though non-Hermitian and non-$PT$-symmetric,  possesses a real spectrum with lower boundedness
and a positive-definite inner product, and that it can be understood as a generalized harmonic oscillator in the sense of $PV$-pseudo-Hermitian quantum theory.

\section{Noncommutative extension}
In the 1930s, Heisenberg~\cite{Hei} proposed a kind of lattice structures of spacetimes, {\em i.e.}, the quantized spacetime now called the
noncommutative spacetime,
in order to overcome the ultraviolet divergence
in quantum field theory. Later Snyder~\cite{Snyder} applied the idea of spacetime noncommutativity
to construct the Lorentz invariant field theory with a small length scale cut-off.
Since the Seiberg-Witten's seminal work~\cite{Seiberg} on describing some low-energy effective theory of open
strings by means of a  noncommutative  gauge theory,
the physics founded on noncommutative spacetimes has been studied intensively, see, for instance, some review articles~\cite{NCQFT}.
As a result, it is quite natural to ask how an $\eta_+$-pseudo-Hermitian Hamiltonian behaves on a
noncommutative space. That is, it is interesting to investigate whether the $\eta_+$-pseudo-Hermitian symmetry,
the real spectrum
and the positive-definite inner product of a non-Hermitian and non-$PT$-symmetric system maintain or not when the system is extened to a noncommutative space.
Incidentally, one of the authors of the present paper established~\cite{YGMCB} a noncommutative theory of chiral bosons and found that
the self-duality that exists in the usual chiral bosons is broken in the noncommutative chiral bosons.

We consider a general two-dimensional canonical noncommutative space with noncommutative spatial coordinate operators and noncommutative
momentum operators as well,
\begin{equation}
[\hat x_j, \hat x_k] = i\theta {\epsilon}_{jk},\qquad
\lbrack\hat p_j, \hat p_k\rbrack = i\tilde{\theta} {\epsilon}_{jk},\qquad
\lbrack\hat x_j, \hat p_k\rbrack = i{\delta}_{jk}, \qquad
j,k=1,2,\label{NCre}
\end{equation}
where $\epsilon_{12}=-\epsilon_{21}=1$, and $\theta$ and $\tilde{\theta}$ independent of coordinate and momentum operators
are real noncommutative parameters
which are much smaller than the Planck constant.
Therefore, we extend our system (eq.~(\ref{H1})) to this
noncommutative space in a straightforward way,
\begin{equation}
\hat{H}=\frac1 2\left(\hat p_1^2+\hat x_1^2\right)+\frac1 2\left(\hat p_2^2+\hat x_2^2\right)
+i\left[A\left(\hat x_1+\hat x_2\right)+B\left({\hat p_1}+{\hat p_2}\right)\right].\label{ncH1}
\end{equation}
In accordance with the commutation relations in the two spaces, {\em i.e.},  the standard Heisenberg commutation relations and eq.~(\ref{NCre}),
we establish the following relationship between the commutative and noncommutative spaces
up to the first order in $\theta$ and $\tilde{\theta}$,
\begin{equation}
\hat x_j = x_j-\frac{1}{2}\theta\epsilon_{jk}p_k,\qquad
\hat p_j = p_j+\frac{1}{2}\tilde{\theta}\epsilon_{jk}x_k,\label{CNC}
\end{equation}
where  the repeated subscripts mean  summation,
and then rewrite
eq.~(\ref{ncH1}) in terms of the coordinate and momentum operators of the commutative space
still up to the first order in $\theta$ and $\tilde{\theta}$,
\begin{eqnarray}
\mathcal{H} &=&\frac1 2\left(p_1^2+x_1^2\right)+\frac1 2\left(p_2^2+x_2^2\right)+i\left[A\left(x_1+x_2\right)+B\left({p_1}+{p_2}\right)\right]\nonumber\\
& & +\frac{1}{2}(\theta+\tilde{\theta})\left(x_2p_1-x_1p_2\right)-i\left[\frac{1}{2}B\tilde{\theta}\left(x_1-x_2\right)
-\frac{1}{2}A{\theta}\left(p_1-p_2\right)\right].\label{ncH2}
\end{eqnarray}
The last two terms in the above Hamiltonian give the noncommutative corrections, where the first that presents the coupling of the two one-dimensional
oscillators is Hermitian while the second is not.
Note that this Hamiltonian is still non-Hermitian and non-$PT$-symmetric, and  that it is also coupled and non-diagonalized.
In addition, we point out that eq.~(\ref{ncH1}) is symmetric under the permutation of
dimension ${}^{\sharp} 1$ and dimension ${}^{\sharp} 2$
while eq.~(\ref{ncH2}) does not possess such a permutation symmetry because
the relationship between the commutative and noncommutative spaces (eq.~(\ref{CNC})) breaks this symmetry under the first order approximation to
the noncommutative parameters.

After analyzing eq.~(\ref{ncH2}), 
we partially diagonalize it up to the first order in $\theta$ and $\tilde{\theta}$,
\begin{eqnarray}
\mathcal{H} = \frac{1}{2}(\mathcal{P}^2_1+\mathcal{X}^2_1)+\frac{1}{2}(\mathcal{P}^2_2+\mathcal{X}^2_2)
 +\frac{1}{2}(\theta+\tilde\theta)(\mathcal{X}_2\mathcal{P}_1-\mathcal{X}_1\mathcal{P}_2)+(A^2+B^2),\label{ncH3}
\end{eqnarray}
where the new variables 
are defined as follows:
\begin{eqnarray}
\mathcal{P}_1\equiv p_1+i\mathcal{B}_1, && \mathcal{X}_1\equiv x_1+i\mathcal{A}_1,\nonumber\\
\mathcal{P}_2\equiv p_2+i\mathcal{B}_2, && \mathcal{X}_2\equiv x_2+i\mathcal{A}_2,\label{nco}
\end{eqnarray}
and new real parameters $\mathcal{A}_j$ and $\mathcal{B}_j$, $j=1,2$, are defined by
\begin{eqnarray}
\mathcal{A}_1\equiv A+\frac{1}{2}B{\theta}, && \mathcal{A}_2\equiv A-\frac{1}{2}B{\theta},\nonumber\\
\mathcal{B}_1\equiv B-\frac{1}{2}A\tilde{\theta}, && \mathcal{B}_2\equiv B+\frac{1}{2}A\tilde{\theta}.
\end{eqnarray}
The new variables, $\mathcal{X}_j$ and $\mathcal{P}_j$, where $j=1,2$, are non-Hermitian like that in the commutative case (see eq.~(\ref{NewV})), and satisfy the same Heisenberg  commutation relations as eq.~(\ref{cr2}),
which is crucial for us to apply our method to the noncommutative extension.

We note that the third term in eq.~(\ref{ncH3}) gives the first order correction in the noncommutative parameters.
This term describes the coupling of oscillator
${}^{\sharp} 1$ and oscillator  ${}^{\sharp} 2$ and thus needs to be dealt with particularly because no couplings exist in the commutative case.

Following the procedure stated in the above section for searching for the operator $V$,
we at first find out the corresponding operator $\mathcal{V}$ for the noncommutative case,
$\mathcal{V}\equiv \mathcal{V}_1\mathcal{V}_2$, where $\mathcal{V}_1\equiv (-1)^{\mathcal{H}_1-\frac{1}{2}}$ and  $\mathcal{V}_2\equiv (-1)^{\mathcal{H}_2-\frac{1}{2}}$, 
and ${\mathcal{H}}_j$'s are defined as
${\mathcal{H}}_j\equiv \frac{1}{2}(\mathcal{P}^2_j+\mathcal{X}^2_j)$, $j=1,2$. 
Note that $\mathcal{V}$ is linear and invertible, but non-Hermitian, and it
is also $P$-pseudo-Hermitian self-adjoint as $V$, {\em i.e.}, $P^{-1}\mathcal{V}^\dagger P=\mathcal{V}$, whose verification is similar to  eq.~(\ref{VPseudo}), see Appendix B. Then we give the expected operator $\eta_+$ as 
$\eta_+=P\mathcal{V}$, 
which can be proved to be Hermitian though $\mathcal{V}$ is not.


Although it is not easy to set a suitable $\boldsymbol{a}_j$ 
because the coupling term appears in the noncommutative case (see eq.~(\ref{ncH3})),
we find out the desired
$\boldsymbol{a}_j$,
\begin{eqnarray}
\boldsymbol{a}_j = \frac{1}{2}\Big((\eta_{jk}+i{\epsilon}_{jk})\mathcal{X}_k+(i\eta_{jk}-{\epsilon}_{jk})\mathcal{P}_k\Big),
\end{eqnarray}
where  the repeated subscripts denote  summation, and $\eta_{jk}\equiv {\rm diag}(1, -1)$.  By considering the $P\mathcal{V}$-pseudo-Hermiticity of $\mathcal{X}_j$ and $\mathcal{P}_j$, 
which is same as that in the commutative case (see eq.~(\ref{XPetaP}) and Appendix C), we therefore
obtain 
from eq.~(\ref{ceation}) the $P\mathcal{V}$-pseudo-Hermitian adjoint to $\boldsymbol{a}_j$,
\begin{eqnarray}
\boldsymbol{a}_j^{\ddagger} = \frac{1}{2}\Big((\eta_{jk}-i{\epsilon}_{jk})\mathcal{X}_k-(i\eta_{jk}+{\epsilon}_{jk})\mathcal{P}_k\Big),
\end{eqnarray}
where  the repeated subscripts denote  summation.
We can show that the algebraic relations of $\boldsymbol{a}_j$ and $\boldsymbol{a}_j^{\ddagger}$ are same as that of the commutative case,
which is the result we expect to.
Further, we give the number operator which is $P\mathcal{V}$-pseudo-Hermitian self-adjoint,
$\mathcal{N}_j= \boldsymbol{a}_j^{\ddagger}\boldsymbol{a}_j$, $j=1,2$,
and find that $\mathcal{N}_j$, $\boldsymbol{a}_j$ and $\boldsymbol{a}_j^{\ddagger}$ have the same commutation relations as that of the commutative case.
Similarly, for a given set of eigenstates of the number operator $\mathcal{N}_j$, {\em i.e.}, $\mathcal{N}_j|n_j\rangle=n_j|n_j\rangle$,
we can prove (see below) that $\boldsymbol{a}_j^{\ddagger}$ and $\boldsymbol{a}_j$ are indeed the creation and annihilation operators we are looking for,
that is, they satisfy the property of ladder operators.

Now we can write the Hamiltonian eq.~(\ref{ncH3}) in a completely diagonalized form by means of the number operator $\mathcal{N}_i$,
\begin{eqnarray}
\mathcal{H} = ({\mathcal{N}}_1+{\mathcal{N}}_2+1)+\frac{1}{2}(\theta+\tilde\theta)({\mathcal{N}}_1-{\mathcal{N}}_2)+(A^2+B^2),\label{ncH5}
\end{eqnarray}
and easily give the real and positive energy spectrum up to the first order in the noncommutative parameters,
\begin{eqnarray}
{\mathcal E}_{n_1n_2} = (n_1+n_2+1)+\frac{1}{2}(\theta+\tilde\theta)(n_1-n_2)+(A^2+B^2),
\end{eqnarray}
where   $n_1, n_2=0, 1, 2, \cdots$.
Note that the first order correction of the spectrum is proportional to the difference between the eigenvalue of oscillator ${}^{\sharp} 1$ and
that of oscillator ${}^{\sharp} 2$.
We point out
that the first order correction of the energy spectrum is vanishing when the noncommutative parameters satisfy the special relation
$\theta+\tilde\theta=0$,
in which case higher order corrections might be considered.
Moreover, if $\theta+\tilde\theta\neq0$ but $n_1- n_2=0$, {\em i.e.}, the energy eigenvalues of oscillator ${}^{\sharp} 1$ and
oscillator ${}^{\sharp} 2$  equal,
there is no first order correction for the spectrum, either. For instance,
it is obvious that the energy level of the ground state is not modified
because of  $n_1= n_2=0$.
However, we emphasize that the noncommutative corrections of the eigenfunction are
non-vanishing even for the two cases ($\theta+\tilde\theta=0$, and $\theta+\tilde\theta\neq0$ but $n_1- n_2=0$)
because the eigenfunction, as stated in the above section, has the same formulation (see the next paragraph for a detailed analysis)
as eqs.~(\ref{ES}) and (\ref{ES1}) with the replacement of ${X}_j$
by the new coordinates $\mathcal{X}_j$  ($j=1,2$) given in eq.~(\ref{nco}), and thus contains the noncommutative parameter
$\theta$ through $\mathcal{X}_j$. This would be seen more evidently from eq.~(\ref{ncH5}) which is the diagonalized form of eq.~(\ref{ncH3}).

We turn to the proof of the positive-definite inner product in the noncommutative case, which shows as in the commutative case that
$\boldsymbol{a}_j^{\ddagger}$ and $\boldsymbol{a}_j$ are the creation and annihilation operators that satisfy
the property of ladder operators.
Because the coupling part is commutative with the free part in the Hamiltonian eq.~(\ref{ncH3}),
that is,\footnote{Such a commutativity can be seen more clearly from eq.~(\ref{ncH5}), {\em i.e.},
$[{\mathcal{N}}_1-{\mathcal{N}}_2, {\mathcal{N}}_1+{\mathcal{N}}_2+1]=0$.}
\begin{equation}
\left[\mathcal{X}_2\mathcal{P}_1-\mathcal{X}_1\mathcal{P}_2,
\frac{1}{2}(\mathcal{P}^2_1+\mathcal{X}^2_1)+\frac{1}{2}(\mathcal{P}^2_2+\mathcal{X}^2_2)\right]=0,
\end{equation}
we conclude that the eigenfunction of the total Hamiltonian (eq.~(\ref{ncH3})) is the product of the eigenfunctions of the oscillator
${}^{\sharp} 1$ Hamiltonian and oscillator  ${}^{\sharp} 2$
Hamiltonian. As a result,
it takes the same form as that obtained in the above section just with the replacement of
$X_j$ by $\mathcal{X}_j$, where $j=1,2$. 
For example, the eigenfunction of the ground state
for one of the oscillators is:
$\varphi_0(\mathcal{X}_j)= \frac{1}{\sqrt[4]{\pi}}\exp(-\frac{1}{2}\mathcal{X}_j^2+\mathcal{B}_j\mathcal{X}_j)$, where $j=1,2$, and repeated subscripts
do not sum.
Similar to the commutative case in section 3
(see eq.~(\ref{PVinner})),
by using $\mathcal{V}|0\rangle =(-1)^{\mathcal{H}_1+\mathcal{H}_2-1} |0\rangle=(-1)^{\mathcal{N}_1+\mathcal{N}_2} |0\rangle=|0\rangle$,
we have $\langle0|P\mathcal{V}|0\rangle =\langle0|P|0\rangle$, and can therefore prove
the normalization of the ground state under the generalized inner product  in terms of the Cauchy's
residue theorem together with the contour chosen in Figure 1,
{\em i.e.}, $\langle0|P\mathcal{V}|0\rangle =\langle0|P|0\rangle=1$. Moreover, we can also prove the orthogonality and normalization of the inner products
of excited states, like eq.~(\ref{PVinner1}) for the noncommutative case.
This completes the proof of the positive definiteness of the
generalized inner product defined by eq.~(\ref{innerpro}) with $\eta_+=P\mathcal{V}$.

As analyzed in section 3 for the commutative case, we can verify straightforwardly from eq.~(\ref{ncH5}) that the Hamiltonian inherently has the
$P\mathcal{V}$-pseudo-Hermiticity in the noncommutative case,
\begin{equation}
\mathcal{H}^{\ddagger} \equiv (P\mathcal{V})^{-1}\mathcal{H}^{\dagger}(P\mathcal{V})=\mathcal{H},\label{calVPseudo}
\end{equation}
which was unknown before the finding of $\eta_+=P\mathcal{V}$.
 As a consequence,
in the noncommutative generalization we confirm that the reality of energy spectra with lower boundedness
and the positive definiteness of inner products 
maintain
because the $P\mathcal{V}$-pseudo-Hermitian symmetry exists in the system depicted by the Hamiltonian eq.~(\ref{ncH2}), or eq.~(\ref{ncH3}),
or eq.~(\ref{ncH5}).
In addition, besides this Hamiltonian, the other physical observables whose average values are real also have $P\mathcal{V}$-pseudo-Hermiticity, such as the coordinate $\mathcal{X}_j$ and the momentum $\mathcal{P}_j$.

Following the above treatment for investigating the effect of the first order correction on the noncommutative model, we can calculate the higher order corrections, such as the second order correction. We shall see an interesting property that the second order correction of the Hamiltonian is decoupled and thus the treatment 
for the first order correction is still applicable to the second order case. The details are given below.

At first we propose the following relationship between the commutative and noncommutative spaces
up to the second order in $\theta$ and $\tilde{\theta}$,
\begin{equation}
\hat x_j = \left(1-\frac{1}{8}\theta\tilde{\theta}\right)x_j-\frac{1}{2}\theta\epsilon_{jk}p_k,\qquad
\hat p_j = \left(1-\frac{1}{8}\theta\tilde{\theta}\right)p_j+\frac{1}{2}\tilde{\theta}\epsilon_{jk}x_k,\label{2ndCNC}
\end{equation}
where the repeated subscripts mean summation. Substituting the above relations into eq. (\ref{ncH1}), we obtain the Hamiltonian 
in terms of the coordinate and momentum operators of the commutative space
still up to the second order in $\theta$ and $\tilde{\theta}$,
\begin{eqnarray}
\mathcal{H^{\prime}} &=&\frac1 2\left(p_1^2+x_1^2\right)+\frac1 2\left(p_2^2+x_2^2\right)+i\left[A\left(x_1+x_2\right)+B\left({p_1}+{p_2}\right)\right]\nonumber\\
& & +\frac{1}{2}(\theta+\tilde{\theta})\left(x_2p_1-x_1p_2\right)-i\left[\frac{1}{2}B\tilde{\theta}\left(x_1-x_2\right)
-\frac{1}{2}A{\theta}\left(p_1-p_2\right)\right] \nonumber \\
& &+ \frac{1}{2}\left[\frac{\theta(\theta-\tilde{\theta})}{4}\left(p_1^2+p_2^2\right)+\frac{\tilde{\theta}(\tilde{\theta}-\theta)}{4}\left(x_1^2+x_2^2\right)\right]\nonumber \\
& & -i \frac{1}{8}\theta\tilde{\theta}\left[A\left(x_1+x_2\right)+B\left({p_1}+{p_2}\right)\right], \label{2ndncH2}
\end{eqnarray}
where the first  two lines of the above equation exactly cover eq. (\ref{ncH2}), {\em i.e.},  the Hamiltonian up to the first order in $\theta$ and $\tilde{\theta}$. One can see that the last two lines give the second order correction and they are decoupled. 

Then, following the treatment given for $\mathcal{H}$ (eq.~(\ref{ncH2})), we partially diagonalize $\mathcal{H^{\prime}}$ up to the second order in $\theta$ and $\tilde{\theta}$,
\begin{eqnarray}
\mathcal{H^{\prime}} &= & \frac12({{\mathcal P}^{\prime}_1}^2+{{\mathcal X}^{\prime}_1}^2)+ \frac12({{\mathcal P}^{\prime}_2}^2+{{\mathcal X}^{\prime}_2}^2)
+\frac12(\theta+\tilde{\theta})({{\mathcal X}^{\prime}_2}{{\mathcal P}^{\prime}_1}-{{\mathcal X}^{\prime}_1}{{\mathcal P}^{\prime}_2}) +(A^2+B^2) \nonumber \\
& &+\frac12
\left[\frac{\theta(\theta-\tilde{\theta})}{4}({{\mathcal P}^{\prime}_1}^2+{{\mathcal P}^{\prime}_2}^2)+
\frac{\tilde{\theta}(\tilde{\theta}-\theta)}{4}({{\mathcal X}^{\prime}_1}^2+{{\mathcal X}^{\prime}_2}^2)\right], \label{2ndncH3}
\end{eqnarray}
where the new symbols are defined as follows:
\begin{eqnarray}
\mathcal{P}^{\prime}_1\equiv p_1+i\mathcal{B}^{\prime}_1, && \mathcal{X}^{\prime}_1\equiv x_1+i\mathcal{A}^{\prime}_1,\nonumber\\
\mathcal{P}^{\prime}_2\equiv p_2+i\mathcal{B}^{\prime}_2, && \mathcal{X}^{\prime}_2\equiv x_2+i\mathcal{A}^{\prime}_2,\label{2ndPp}
\end{eqnarray}
and
\begin{eqnarray}
\mathcal{A}^{\prime}_1\equiv A+\frac{1}{2}B{\theta} + \frac38 A\theta \tilde{\theta}, && \mathcal{A}^{\prime}_2\equiv A-\frac{1}{2}B{\theta}+\frac38 A\theta \tilde{\theta},\nonumber\\
\mathcal{B}^{\prime}_1\equiv B-\frac{1}{2}A\tilde{\theta}+\frac38 B\theta \tilde{\theta}, && \mathcal{B}^{\prime}_2\equiv B+\frac{1}{2}A\tilde{\theta}+\frac38 B\theta \tilde{\theta}.\label{2ndAB}
\end{eqnarray}

Next, considering the second order correction in eq.~(\ref{2ndncH3}) we find the corresponding annihilation and creation operators,
\begin{eqnarray}
{\boldsymbol a}^{\prime}_j &=& \frac{1}{2}\Big((\eta_{jk}+i{\epsilon}_{jk})\sqrt{M\Omega}\mathcal{X}^{\prime}_k+(i\eta_{jk}-{\epsilon}_{jk})\sqrt{\frac{\Omega}{K}}\mathcal{P}^{\prime}_k\Big),
\nonumber \\
{{\boldsymbol a}^{\prime}_j}^{\ddagger} &=& \frac{1}{2}\Big((\eta_{jk}-i{\epsilon}_{jk})\sqrt{M\Omega}\mathcal{X}^{\prime}_k-(i\eta_{jk}+{\epsilon}_{jk})\sqrt{\frac{\Omega}{K}}\mathcal{P}^{\prime}_k\Big),\label{2nda}
\end{eqnarray}
where the new parameters are defined by
\begin{equation}
\frac{1}{M} \equiv 1+\frac{\theta(\theta-\tilde{\theta})}{4}, \qquad
K  \equiv 1+\frac{\tilde{\theta}(\tilde{\theta}-\theta)}{4}, \qquad \Omega \equiv \sqrt{\frac{K}{M}}.\label{2nd3para}
\end{equation}
After introducing the number operator $\mathcal{N}^{\prime}_j$ associated with ${\boldsymbol a}^{\prime}_j$ and ${{\boldsymbol a}^{\prime}_j}^{\ddagger}$ as $\mathcal{N}^{\prime}_j ={{\boldsymbol a}^{\prime}_j}^{\ddagger}{\boldsymbol a}^{\prime}_j$, where $j=1,2$,
we can rewrite the Hamiltonian $\mathcal{H^{\prime}}$ in a completely diagonalized form,
\begin{eqnarray}
\mathcal{H^{\prime}} &=& \frac{1}{2M}({{\mathcal P}^{\prime}_1}^2+{{\mathcal P}^{\prime}_2}^2)+\frac{K}{2}({{\mathcal X}^{\prime}_1}^2+{{\mathcal X}^{\prime}_2}^2)+\frac12(\theta+\tilde{\theta})({{\mathcal X}^{\prime}_2}{{\mathcal P}^{\prime}_1}-{{\mathcal X}^{\prime}_1}{{\mathcal P}^{\prime}_2})+(A^2+B^2) \nonumber \\
& = & \Omega\,(\mathcal{N}^{\prime}_1 +\mathcal{N}^{\prime}_2+1)+\frac12(\theta+\tilde{\theta})(\mathcal{N}^{\prime}_1-\mathcal{N}^{\prime}_2)+(A^2+B^2).\label{2ndncH4}
\end{eqnarray}
As a result, the energy spectrum up to the second order in $\theta$ and $ \tilde{\theta}$ reads
\begin{eqnarray}
{\mathcal E}^{\prime}_{n^{\prime}_1n^{\prime}_2} = \left(1+\frac{1}{8}(\theta-\tilde{\theta})^2\right)(n^{\prime}_1+n^{\prime}_2+1)+\frac{1}{2}(\theta+\tilde\theta)(n^{\prime}_1-n^{\prime}_2)+(A^2+B^2),\label{2ndenergy}
\end{eqnarray}
where $n^{\prime}_1, n^{\prime}_2=0, 1, 2, \cdots$.

As last, we note that ${\boldsymbol a}^{\prime}_j$ and ${{\boldsymbol a}^{\prime}_j}^{\ddagger}$ are $\eta_+^{\prime}$-pseudo-Hermitian adjoint to each other, and  $\mathcal{N}^{\prime}_j$ is  $\eta_+^{\prime}$-pseudo-Hermitian self-adjoint, where $\eta_+^{\prime}=P\mathcal{V}^{\prime}$. Similar to the first order case, here   $\mathcal{V}^{\prime}\equiv {\mathcal V}^{\prime}_1{\mathcal V}^{\prime}_2$,  ${\mathcal V}^{\prime}_1\equiv (-1)^{{\mathcal H}^{\prime}_1-\frac{1}{2}}$ and  ${\mathcal V}^{\prime}_2\equiv (-1)^{{\mathcal H}^{\prime}_2-\frac{1}{2}}$, and in particular, ${\mathcal H}^{\prime}_j\equiv \frac{1}{2M}{{\mathcal P}_j^{\prime}}^2+\frac{K}{2}{{\mathcal X}_j^{\prime}}^2$, where $j=1, 2$. The other related properties can be discussed similarly. We omit them.

In addition, we mention that the perturbation method in non-Hermitian quantum mechanics can only work effectively when the main part and the perturbation part of a Hamiltonian have the same $\eta$-pseudo Hermiticity.  However, this requirement is not usually satisfied, such as in our noncommutative model $\mathcal{H^{\prime}}$ (see eq.~(\ref{2ndncH2})) when the second order correction terms are dealt with as perturbation. Therefore, it is not convenient to apply the perturbation method for non-Hermitian Hamiltonian systems.  As a whole, we may say it is a more fundamental method to realize the diagonalization  for a coupled non-Hermitian Hamiltonian.

At the end of this section,
it is quite evident that the eigenvalues and eigenfunctions
of our noncommutative generalization turn back to their commutative counterparts (see section 3)
when the parameters $\theta$ and $\tilde\theta$ tend to zero. This shows that our noncommutative extension is consistent.

\section{Conclusion}
In this paper, we provide a possible method for a non-Hermitian and non-$PT$-symmetric quantum system. The crucial
points of this method are to find out the  $\eta_+$ (positive-definite metric) operator and then to define the corresponding 
annihilation, creation and number operators as in eqs.~(\ref{ceation}) and (\ref{number}).
After the $\eta_+$-pseudo-Hermiticity that the non-Hermitian and non-$PT$-symmetric Hamiltonian inherently possesses is found, the real spectrum is given and the positive-definite inner product, the probability explanation of wave functions, the orthogonality of eigenstates, and the unitarity of time evolution can be confirmed. We apply our method at first to a decoupled system
and then to a coupled one by extending the former to the canonical noncommutative space
with both noncommutative spatial coordinate operators and noncommutative momentum operators.
For the two systems, we find out the specific  $\eta_+$ operators and
prove the reality of energy spectra and the positive definiteness of inner products, together with
the probability explanation of wave functions, the orthogonality of eigenstates, and the unitarity of time evolution. 
Moreover, to the coupled system we obtain the first and second order
corrections of spectra in the noncommutative parameters.

Our results show that it is not mandatory to adopt the biorthonormal basis~\cite{Wong,Ali1,Ali2,Ali3} for pseudo-Hermitian systems, and that
it is still available to use the usual orthonormal basis
if the operator $\eta_+$ 
is found.
In other words, it is not necessary to introduce $\eta_{+}$ when describing a non-Hermitian Hamiltonian ($H \neq H^\dag$) in terms of the biorthonormal basis because the set of eigenstates of $H$ is orthogonal  to the set of eigenstates of $H^\dag$. However, in our paper the starting point is to apply the usual orthonormal basis to deal with a non-Hermitian Hamiltonian, thus to introduce $\eta_{+}$ is crucial in order to construct a positive-definite and orthogonal inner product (see eq.~(\ref{innerpro})) for such a  non-Hermitian Hamiltonian.
Therefore, we provide a possible method for dealing with non-Hermitian and non-$PT$-symmetric quantum systems, which is complementary to the $PT$-symmetric method.

We note that our two-step method is applicable to the decoupled Hamiltonian given in section 3 and to the coupled Hamiltonian constructed in section 4, too. We recall that the method based on annihilation and creation operators is not applicable to all coupled (and thus non-diagonalized) Hamiltonians even in the ordinary (Hermitian) quantum mechanics. What we can confirm here is that our method as an earliest attempt from the point of view of  annihilation and creation operators is applicable to some coupled and non-diagonalized Hamiltonians in non-Hermitian quantum mechanics. 
In principle, this two-step method is applicable to other more complex systems, such as a many-body system. The prerequisite is that one has to diagonalize the many-body system at first, and then uses our method. 
In fact, even in the ordinary quantum mechanics the Fock space approach is usually effective to a diagonalized Hamiltonian system, such as the harmonic oscillator.  Our method is a generalization of the Fock space approach to non-Hermitian and non-$PT$-symmetric systems. As a result, for a many-body system governed by a non-Hermitian Hamiltonian,  one can still use our two-step method after diagonalizing it.


In addition, we make a comparison between the ordinary (Hermitian) quantum mechanics and the
$\eta_+$-pseudo-Hermitian quantum mechanics.
For the former the definitions of the bra and ket vector states,
of the inner product, and of the annihilation and creation operators, {\em etc.}, are model-independent,
while for the latter the definitions of the relavent quantities highly depend on the symmetric
operator $\eta_+$ of pseudo-Hermitian systems and thus they are model-dependent since
the Hermitian operator $\eta_+$ is usually model-dependent.
Such a difference has been shown obviously in the present paper. It can easily be understood because
the Hermitian quantum systems are
the special case when $\eta_+$
is fixed to be the identity operator from the point of view of $\eta_+$-pseudo-Hermitian quantum systems.
As a consequence, the former is model-independent because the identity is the only symmetric operator for
all Hermitian systems, but the latter
is model-dependent because every pseudo-Hermitian system in general has its own symmetric operator $\eta_+$.
One has to determine the (positive-definite metric) operator $\eta_+$
that is in general different from one model to another in
pseudo-Hermitian quantum mechanics. It is in principal a hard job to find out an $\eta$ operator for every non-Hermitian and non-$PT$-symmetric quantum system,
but after all a possible method suggested in this paper is available.

At last, we note that the operator $\eta_+$ is not unique to a non-Hermitian Hamiltonian. For instance, to the non-Hermitian and non-$PT$-symmetric Hamiltonian we construct in section 3, see eq.~(\ref{H1}), we find another $\eta_+$ (see Appendix D for the detailed derivation),
$\eta_+={\Lambda}^{\dagger}\Lambda$, 
where 
$\Lambda=\exp \left\{-B\left(x_1+x_2\right)+A\left(p_1+p_2\right)\right\}$.
We can verify that the Hamiltonian eq.~(\ref{H1}) satisfies this $\eta_+$-pseudo-Hermitian self-adjoint condition: $H^{\ddagger}\equiv \eta_+^{-1}H^{\dagger}\eta_+=H$.
We emphasize that it is only an alternative way for us to give real eigenvalues through the $\Lambda$  transformation (see Appendix D) for the non-Hermitian and non-$PT$-symmetric model. To obtain real eigenvalues is not the sole job for establishing a consistent non-Hermitian quantum theory. In general, one has to consider the
other indispensable ingredients, such as the positive definite inner product that relates to the probability explanation of wave functions, the orthogonality of eigenstates, and the unitarity of time evolution, etc.  Section 2 of the present paper contains these indispensable ingredients and thus plays an important role in establishing a consistent quantum theory for non-Hermitian Hamiltonians.


\section*{Acknowledgments}
Y-GM would like to thank
H.P. Nilles of the University of Bonn for kind hospitality.
This work was supported in part by the Alexander von Humboldt Foundation under a short term programme, by the National Natural
Science Foundation of China under grants No.11175090 and
by the Ministry of Education of China under grant No.20120031110027.

\newpage

\section*{Appendix A \hspace{.24cm}Derivation of some useful formulae}
We shall use our specific notation of the modified bra vector state (see eq.~(\ref{stateadjoint})) in the derivations below.
It is obvious that this process reduces apparently to that of the
ordinary (Hermitian) quantum mechanics when $\eta_+$ becomes the trivial identity operator.

Let us at first derive the $n$-particle state. If $|0 \rangle$ stands for the ground state and $a$ annihilates the ground state, $a|0 \rangle =0$,
we calculate the average value of $a^n(a^{\ddagger})^n$ with respect to the ground state and
its $\eta_+$-pseudo-Hermitian adjoint (the modified bra vector state, see eq.~(\ref{stateadjoint}))
by repeatedly using eq.~(\ref{acalgrbra}),
\begin{eqnarray}
{}^\ddagger \langle 0|a^n(a^{\ddagger})^n|0\rangle  \equiv  \langle 0|\eta_+a^n(a^{\ddagger})^n|0\rangle = n!\,\langle 0|\eta_+|0\rangle \equiv n! \left({}^\ddagger \langle 0|0\rangle\right).\label{aadageraverage}
\end{eqnarray}
If $|n\rangle$ is defined by
\begin{eqnarray}
|n\rangle\equiv\frac{1}{\sqrt{n!}}(a^{\ddagger})^n|0\rangle, \label{nstate}
\end{eqnarray}
we obtain its $\eta_+$-pseudo-Hermitian adjoint by using  eqs.~(\ref{stateadjoint}) and (\ref{ceation}),
\begin{eqnarray}
{}^\ddagger \langle n| \equiv \langle 0|\frac{1}{\sqrt{n!}}\left((a^{\ddagger})^n\right)^\dagger \eta_+=\langle 0|\frac{1}{\sqrt{n!}}\eta_+a^n
\equiv {}^\ddagger\langle 0|\frac{1}{\sqrt{n!}}a^n. \label{nstatead}
\end{eqnarray}
Therefore, we can rewrite  eq.~(\ref{aadageraverage}) using the notation with hidden $\eta_+$ as
\begin{eqnarray}
{}^\ddagger \langle n|n\rangle={}^\ddagger\langle 0|\frac{1}{n!}a^n(a^{\ddagger})^n|0\rangle={}^\ddagger\langle 0|0\rangle.\label{n0norm}
\end{eqnarray}
Moreover, by using eqs.~(\ref{acalgrbra}), (\ref{number}) and (\ref{nstate}) and considering $a|0 \rangle =0$ again, we derive
\begin{eqnarray}
N |n\rangle \equiv  \frac{1}{\sqrt{n!}} a^\ddagger a (a^{\ddagger})^n|0\rangle=n |n\rangle. \label{Neigenstate}
\end{eqnarray}
Combining  eq.~(\ref{n0norm}) and eq.~(\ref{Neigenstate}), we obtain
\begin{eqnarray}
{}^\ddagger \langle n|N|n\rangle =n \left({}^\ddagger \langle n|n\rangle\right)=n \left( {}^\ddagger\langle 0|0\rangle\right).\label{Naverage}
\end{eqnarray}
Consequently, if the ground state is
normalized with respect to the generalized inner product (eq.~(\ref{innerpro})),
{\em i.e.}, ${}^\ddagger\langle 0|0\rangle \equiv \langle 0|\eta_+|0\rangle=1$, the state
defined by eq.~(\ref{nstate}) is convinced to be the expected  $n$-particle state and the operator $N$ defined by eq.~(\ref{number}) is then confirmed to be the desired number operator. Moreover, we can deduce the orthogonality of eigenstates, ${}^\ddagger \langle n|m\rangle \equiv \langle n|\eta_+|m\rangle=\delta_{nm}$. In 
sections 3 and 4, for instance,  the exact forms of $\eta_+$ operators are provided for the concrete models, the ground state can be determined to be normalized,  and the orthogonality of eigenstates is guaranteed.

Now we calculate the ladder property of redefined creation and annihilation operators. It is straightforward from eq.~(\ref{nstate}) to have
\begin{eqnarray}
{a}^\ddagger |n\rangle \equiv \frac{1}{\sqrt{n!}}  (a^{\ddagger})^{n+1}|0\rangle = \frac{\sqrt{n+1}}{\sqrt{(n+1)!}}  (a^{\ddagger})^{n+1}|0\rangle
=\sqrt{n+1}\, |n+1\rangle.\nonumber
\end{eqnarray}
Multiplying the above equation by the operator $a$ from the left and using eqs.~(\ref{acalgrbra}), (\ref{number}) and (\ref{Neigenstate}),
we get 
\[
a {a}^\ddagger |n\rangle =  ({a}^\ddagger a +1) |n\rangle = (N+1) |n\rangle = (n+1) |n\rangle.
\]
Combining the above two equations, we obtain $ a |n+1\rangle = \sqrt{n+1} |n\rangle$.
As a result, we give the following ladder properties for the operators
${a}^\ddagger$ and $a$, respectively,
\begin{eqnarray}
a^{\ddagger}|n\rangle = \sqrt{n+1}\,|n+1\rangle,\qquad a|n\rangle =\sqrt{n}\,|n-1\rangle,\label{ladder}
\end{eqnarray}
which indeed shows that ${a}^\ddagger$ has the function of creation and $a$ that of annihilation as expected.


\section*{Appendix B \hspace{.24cm}Verification of eq.~(\ref{VPseudo})}
Using eq.~(\ref{V}) and the Taylor expansion, we have
\begin{eqnarray}
& &P^{-1}V^{\dagger}P\nonumber \\
&=& P^{-1}(-1)^{H_1^{\dagger}+H_2^{\dagger}-1}P \nonumber \\
&=& \sum_{n=0}^{\infty}\frac{\ln^n (-1)}{n!}P^{-1}\left(H_1^{\dagger}+H_2^{\dagger}-1\right)^nP \nonumber \\
& =&\sum_{n=0}^{\infty}\frac{\ln^n (-1)}{n!}\underbrace{\left\{P^{-1}\left(H_1^{\dagger}+H_2^{\dagger}-1\right)P\right\}  \cdots \left\{P^{-1}\left(H_1^{\dagger}+H_2^{\dagger}-1\right)P\right\}}_{\text{n\,terms}}. \label{v3}
\end{eqnarray}
Considering eq.~(\ref{NewV}), we get
\begin{equation}
P^{-1}\left(H_1^{\dagger}+H_2^{\dagger}-1\right)P=H_1+H_2-1.\label{v4}
\end{equation} 
Substituting eq.~(\ref{v4}) into eq.~(\ref{v3}), we thus obtain
\begin{eqnarray}
P^{-1}V^{\dagger}P=\sum_{n=0}^{\infty}\frac{\ln^n (-1)}{n!}\left(H_1+H_2-1\right)^n= (-1)^{H_1+H_2-1}=V.\label{v5}
\end{eqnarray}

\section*{Appendix C \hspace{.24cm}Verification of eq.~(\ref{XPetaP})}
We just verify the case $j=1$, {\em i.e.}, $X_1^{\ddagger} \equiv  \left(PV\right)^{-1}X_1^{\dag}\left(PV\right)= X_1$ and $P_1^{\ddagger} \equiv  \left(PV\right)^{-1}P_1^{\dag}\left(PV\right)= P_1$. As to $j=2$, the procedure is exactly the same.

Starting from
\begin{eqnarray} 
\left(PV\right)^{-1}X_1^{\dag}\left(PV\right) &=&V^{-1}\left(P^{-1}X_1^{\dag}P\right)V,\nonumber \\
\left(PV\right)^{-1}P_1^{\dag}\left(PV\right) &=&V^{-1}\left(P^{-1}P_1^{\dag}P\right)V,\nonumber
\end{eqnarray}
we at first get $P^{-1}X_1^{\dag}P=-X_1$ and $P^{-1}P_1^{\dag}P=-P_1$ in terms of eq.~(\ref{NewV}). Then, considering eq.~(\ref{cr2}) and eq.~(\ref{V}) we obtain
\begin{eqnarray}
\left(PV\right)^{-1}X_1^{\dag}\left(PV\right)&=&-V^{-1}X_1V=-V_1^{-1}X_1V_1, \label{v6}\\
\left(PV\right)^{-1}P_1^{\dag}\left(PV\right)&=&-V^{-1}P_1V=-V_1^{-1}P_1V_1.\label{v7}
\end{eqnarray}

When applying the BCH formula,
\begin{eqnarray}
e^{-D}Ce^{D}
=C+[C, D]+\frac{1}{2!}[[C, D], D]+\frac{1}{3!}[[[C, D], D], D]+\cdots,\label{BCH}
\end{eqnarray}
to eq.~(\ref{v6}), {\em i.e.}, letting $V_1^{-1}X_1V_1=e^{-D}Ce^{D}$, we have the corresponding operators $C$ and $D$ with eq.~(\ref{NewV}) and eq.~(\ref{V}),
\begin{eqnarray} 
C &=&X_1,\label{CX} \\
D &=&\frac{\ln(-1)}{2} \left(P_1^2+X_1^2-1\right).\label{DX}
\end{eqnarray}
Thus, we compute the commutation relations by using eq.~(\ref{cr2}),
\begin{eqnarray}
\begin{split}
[C, D]  &&=&& i \ln(-1)P_1,\nonumber \\
[[C, D], D]  &&=&& \ln^2 (-1)X_1,\nonumber \\
[[[C, D], D], D]  &&=&& i \ln^3 (-1)P_1,\nonumber \\
[[[[C, D], D], D], D]  &&=&&  \ln^4 (-1)X_1,\\
 &&\vdots&& \nonumber
\end{split}
\end{eqnarray}
and acquire
\begin{eqnarray} 
V_1^{-1}X_1V_1=X_1+i \ln(-1)P_1+\frac{1}{2!}\ln^2 (-1)X_1+\frac{1}{3!}i \ln^3 (-1)P_1+\frac{1}{4!} \ln^4 (-1)X_1+\cdots.\label{X1}
\end{eqnarray}
Further considering
\begin{eqnarray}
\ln(-1)=i(2k+1)\pi, \qquad
k=0, \pm{1}, \pm{2}, \cdots, \nonumber
\end{eqnarray}
we have
\begin{eqnarray} 
V_1^{-1}X_1V_1&=&\sum_{n=0}^{\infty}\frac{(-1)^n}{(2n)!}\{(2k+1)\pi\}^{2n}X_1-\sum_{n=0}^{\infty}\frac{(-1)^n}{(2n+1)!}\{(2k+1)\pi\}^{2n+1}P_1,\nonumber \\
& =&\cos \{(2k+1)\pi\}X_1-\sin \{(2k+1)\pi\}P_1,\nonumber \\
&=&-X_1.\label{X2}
\end{eqnarray}
Combining eq.~(\ref{v6}) with eq.~(\ref{X2}), we reach our goal,
\begin{eqnarray}
\left(PV\right)^{-1}X_1^{\dag}\left(PV\right)=X_1.
\end{eqnarray}

In addition, when we apply the BCH formula to eq.~(\ref{v7}), {\em i.e.}, let $V_1^{-1}P_1V_1=e^{-{\tilde D}}{\tilde C}e^{{\tilde D}}$, the corresponding operator ${\tilde C}$ takes the form,
\begin{eqnarray} 
{\tilde C} =P_1,\label{CP}
\end{eqnarray}
and ${\tilde D}$ is same as $D$ (see eq.~(\ref{DX})). Therefore, from the commutation relations,
\begin{eqnarray}
\begin{split}
[{\tilde C}, {\tilde D}]  &&=&& -i \ln(-1)X_1,\nonumber \\
[[{\tilde C}, {\tilde D}], {\tilde D}]  &&=&& \ln^2 (-1)P_1,\nonumber \\
[[[{\tilde C}, {\tilde D}], {\tilde D}], {\tilde D}]  &&=&& -i \ln^3 (-1)X_1,\nonumber \\
[[[[{\tilde C}, {\tilde D}], {\tilde D}], {\tilde D}], {\tilde D}]  &&=&&  \ln^4 (-1)P_1,\\
 &&\vdots&& \nonumber
\end{split}
\end{eqnarray}
we obtain
\begin{eqnarray} 
V_1^{-1}P_1V_1&=&P_1-i \ln(-1)X_1+\frac{1}{2!}\ln^2 (-1)P_1-\frac{1}{3!}i \ln^3 (-1)X_1+\frac{1}{4!} \ln^4 (-1)P_1+\cdots\nonumber \\
&=&\sum_{n=0}^{\infty}\frac{(-1)^n}{(2n)!}\{(2k+1)\pi\}^{2n}P_1+\sum_{n=0}^{\infty}\frac{(-1)^n}{(2n+1)!}\{(2k+1)\pi\}^{2n+1}X_1,\nonumber \\
& =&\cos \{(2k+1)\pi\}P_1+\sin \{(2k+1)\pi\}X_1,\nonumber \\
&=&-P_1,\label{P2}
\end{eqnarray}
and thus verify
\begin{eqnarray}
\left(PV\right)^{-1}P_1^{\dag}\left(PV\right)=-V_1^{-1}P_1V_1=P_1.
\end{eqnarray}

\section*{Appendix D \hspace{.24cm}Derivation of another $\eta_+$}
Suppose $h$ is a Hermitian Hamiltonian that has the same eigenvalues as $H$ (see eq.~(\ref{H1})), but different eigenfunctions,
\begin{eqnarray}
H\Phi=E\Phi, \qquad h\phi=E\phi, \label{Hheigen}
\end{eqnarray} 
one can introduce an operator $\Lambda$, $\Lambda \equiv e^{-{Q}/{2}}$, to connect the two representations,
\begin{eqnarray}
h=\Lambda H {\Lambda}^{-1}=e^{-{Q}/{2}} H e^{{Q}/{2}}, \qquad
\phi = \Lambda \Phi=e^{-{Q}/{2}} \Phi.\label{connef}
\end{eqnarray}  
By using the Baker-Campbell-Hausdorff formula, 
\begin{eqnarray}
h=H+\frac{1}{2}[H,Q]+\frac{1}{2!2^2}[[H,Q],Q]+\frac{1}{3!2^3}[[[H,Q],Q],Q]+\cdots,
\end{eqnarray}
and considering the standard Heisenberg commutation relations of $x_j$ and $p_j$, where $j=1,2$,  and the Hermiticity of $h$, one obtains
\begin{eqnarray}
Q=2B(x_1+x_2) -2A(p_1+p_2),
\end{eqnarray}
and
\begin{eqnarray}
h=\frac{1}{2}\left(p_1^2+p_2^2\right)+\frac{1}{2}\left(x_1^2+x_2^2\right)+(A^2+B^2).
\end{eqnarray}

Now using eq.~(\ref{connef}) and the Hermiticity of $h$, $h^{\dagger}=h$, one deduces
\begin{eqnarray}
H=\left({\Lambda}^{\dagger}\Lambda\right)^{-1}H^{\dagger}\left({\Lambda}^{\dagger}\Lambda\right),
\end{eqnarray}
which gives the $\eta_+$-pseudo-Hermiticity of $H$, where $\eta_+={\Lambda}^{\dagger}\Lambda=e^{-Q}$.

We note that the orthogonality of eigenstates is ensured in both the Hermitian and $\eta_+$-pseudo-Hermitian representations, {\em i.e.},  if $\langle \phi_i|\phi_j\rangle=\delta_{ij}$ in the Hermitian representation, then we certainly deduce  the orthogonality of eigenstates in the $\eta_+$-pseudo-Hermitian representation,
${}^\ddagger\langle \Phi_i|\Phi_j\rangle\equiv \langle \Phi_i|\eta_+|\Phi_j\rangle = \langle \Phi_i|{\Lambda}^{\dagger}\Lambda|\Phi_j\rangle =\langle \phi_i|\phi_j\rangle=\delta_{ij}$.


\end{document}